# Intelligent Automated Diagnosis of Client Device Bottlenecks in Private Clouds

Chathuranga Widanapathirana    Jonathan Li    Y. Ahmet Şekercioğlu
Milosh Ivanovich    Paul Fitzpatrick
Department of Electrical and Computer Systems Engineering, Monash University, Australia

*Abstract*—We present an automated solution for rapid diagnosis of client device problems in private cloud environments: the Intelligent Automated Client Diagnostic (IACD) system. Clients are diagnosed with the aid of Transmission Control Protocol (TCP) packet traces, by (i) observation of anomalous artifacts occurring as a result of each fault and (ii) subsequent use of the inference capabilities of soft-margin Support Vector Machine (SVM) classifiers. The IACD system features a modular design and is extendible to new faults, with detection capability unaffected by the TCP variant used at the client. Experimental evaluation of the IACD system in a controlled environment demonstrated an overall diagnostic accuracy of 98%.

## I. Introduction

With the rapid adoption of the "cloud applications" by the ordinary, non-expert users, "always available" network connections, and consistently fast communication speeds are becoming critically important. The networking research community has converged on the common understanding that performance unpredictability and data transfer bottlenecks are going to be significant obstacles for satisfactory cloud computing experience [1], [2]. A number of recent industry surveys cite that a primary challenge in managing cloud services is the lack of tools to identify the sources of performance bottlenecks and rapid troubleshooting. Additionally, automation is a mandatory requirement for achieving highly scalable cloud services, which present a significant research challenge for creating comprehensive diagnostic solutions [3].

Since, data centers employ experienced technical staff that continuously monitor and resolve performance issues within their systems that host the cloud, connection performance problems experienced by a user are most likely to occur in the client device itself [4]. The problems are often

- the result of overly conservative default networking parameters supplied with the operating system,
- client device configuration errors,
- mismatch between the client device and link settings, or
- could even be some subtle, intermittent hardware errors.

As an example, a relatively large number of operating system parameters can be optimally set leading to significant improvement on user experience, but in practice these settings are difficult for ordinary users to manipulate. Some studies have found that network data rates reached by average users are only one third of those achieved by expert users, a phenomenon commonly referred to as the "Wizard Gap" [5]. Hence, a fully automated and comprehensive solution is required for rapid diagnosis of client device problems.

Diagnosis methods based on collected packet traces over TCP (Transmission Control Protocol) sessions have been shown to be effective for finding the root causes of a number of complicated network problems in special cases [6]. The packet traces contain artifacts that reveal the behavioral characteristics of the underlying network elements, and skilled investigators can use them to pinpoint the location and the root causes of the faults. A great advantage of TCP trace analysis methods is, the trace collection can easily be done quickly and without any special equipment. However, the expertise and resources required for these methods hinder their general applicability. In addition, TCP based diagnosis methods available today do not offer fully automated solutions, easy extendability through learning capabilities, robustness against TCP variants, or ease of implementation [7]–[11].

We think that TCP trace analysis based diagnosis tools can be made fully automated through machine learning methods, and their capabilities can be greatly extended. In this paper, we demonstrate the potential of machine learning by presenting our intelligent inference method "Intelligent Automated Client Diagnostic (IACD)" system, and early results of a feasibility study. In the feasibility study, we focus only on the private cloud[1] environments becaouse they offer relative simplicity in access network architecture as compared to much more complex public or hybrid cloud networks. We will use this work as a starting point towards the creation of a broadly applicable general solution.

The IACD system (i) relies only on collection of packet traces upon reporting of a problem, and (ii) focuses on identifying client device faults and misconfigurations to complement the existing diagnosis capabilities used by the network support personnel.

The paper first presents the overall framework of the IACD system. Then the paper describes the details of the classifier training phase and how the diagnosis is performed. Next, the system performance is evaluated against various client faults and TCP types followed by a discussion on the system characteristics and comparison with other available solutions. Finally, the paper concludes with a summary and a description of the future directions.

---
[1]A "private cloud" (also called an "internal cloud" or "enterprise cloud") is a data center architecture that provides hosted services to a limited number of users behind a firewall [12].

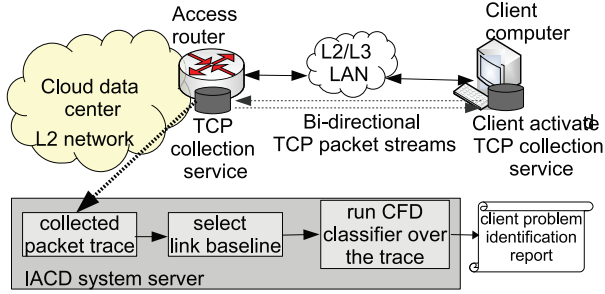

Fig. 1. Overview of the operation of the IACD system

## II. IACD Framework

The outline of the IACD system is shown in Figure 1. The system's input is a TCP packet trace of a known stream of data between the client device and the access router of the data center. The packet trace can easily be obtained through a user initiated process by activating a program on a web page specially created for this purpose.

The machine learning capabilities of the IACD system are provided by the Client Fault Diagnostic (CFD) classifier (Figure 2). The CFD classifier identifies the presence of artifacts belonging to specific types of faults that cause performance problems, and should be capable of both single fault and multiple simultaneous fault classifications. This requirement has bearing on the design choice for selecting either a binary or multi-class classifier. There are studies [13], [14] that compare these approaches for different applications outlining the benefits of each. In our design, we opted to use a parallel network of binary classifier modules (CF-classifiers), each trained to diagnose a single class of fault (Figure 2). This arrangement collectively performs a Multi-class classification. The reasons for this choice are as follows:

- flexibility to continually add new diagnostic capabilities, without the need to retrain the complete system,
- freedom to select classifier parameters optimized to detect a specific type of artifact independent of other fault classifiers,
- parallelism which can shorten classification time in a scalable manner as the number of modules increases.

During the training phase: (i) packet traces are obtained under controlled link conditions emulating healthy and faulty clients (ii) "class labels" are assigned to each trace, (iii) features are extracted, and (iv) feature vector-class label pairs are stored in a database

$$\Theta_{\text{cfd}} = \{(\mathbf{x}_1, y_1), \ldots, (\mathbf{x}_n, y_n)\}. \qquad (1)$$

Here $\mathbf{x}_i \in \Re^m$ is the $m$-dimensional feature vector, $y_i \in \{\text{cf}_0, \ldots, \text{cf}_p\}$ is the class label associated with the feature vector $\mathbf{x}_i$, $p$ is the number of classes, $i = 1, \ldots, n$, and $n$ is the number of packet traces collected. The class label $\text{cf}_0$ is used for a healthy client and $\text{cf}_1, \text{cf}_2, \ldots, \text{cf}_p$ for $p$ types of different client faults. The feature vector $\mathbf{x}_i$, combined with the class label $y_i$, is called the "signature" of the $i^{th}$ instance.

Each CF-classifier module then selects a training data subset ($\Theta_{\text{cf}}^j$) with signatures labeled as $\text{cf}_j$ for the faulty class and $\text{cf}_0$ for the healthy class for training the $j^{th}$ binary CF-classifier as in (3)

$$\Theta_{\text{cf}}^j = \{(\mathbf{x}_k, \text{cf}_0), \ldots, (\mathbf{x}_j, \text{cf}_j)\}, \qquad (2)$$

where $k$ and $j$ constitute the signature subset used in training.

Then, each module independently processes data and selects the unique feature subset (artifact) that separates the two classes. Subsequently, this feature subset is sent to the pattern classifier module to model the classifier boundaries. Each CF-classifier module uses an L2 soft-margin Support Vector Machine (SVM) [15] for pattern classification (we conducted experiments with Decision Trees [16], Artificial Neural Networks [17], Naïve Bayes (NB) [18], k-Nearest Neighbour [19] as well, and found that SVM, so far, yields the most accurate results). During the diagnostic phase, the trained CF-classifiers determine if a trace sample contains the artifacts from each particular fault.

## III. Classifier Training

### A. Data Collection

Training samples are collected either from a test bed to emulate the faults in a well-regulated network or from the live cloud network. The diagnostic accuracy of the classifier is highly dependent on the effectiveness of emulating the fault correctly, and the consistency of the artifacts collected. Using standard packet capture libraries, we collect two traces, one at the client and one at the server. Both traces are captured with bi-directional packet flows with file of size 100MB to ensure most connection details are intercepted.

### B. Trace Signature Creation

The two collected packet traces are analyzed individually, and their extracted features are combined to form an $m$-dimensional feature vector, $\mathbf{x}$ (in equation (1)). We have developed a signature extraction technique, based on an open source trace analysis tool "tcptrace" [20], that extracts 140 different statistical parameters for each trace which forms a combined total of 280 parameters for each signature. For example, a raw feature vector $\mathbf{x}_i$ from a faulty link before any data pre-processing may look like

$$\mathbf{x}_i = \{249256, 295, 0, 32, 39, 1, 1, ...\}, \quad (3)$$

with each numerical value representing an aggregated statistic from the trace such as *total_pkts*, *ack_pkts*, *resets*, *rexmt_data_pkts*, *zero_window_probe_pkts*. This statistical trace characterization technique maps a packet stream into a set of numerical values encapsulating the connection characteristics and preserving the fault artifacts. The signatures are unique, even within the same class due to the minor variations of the connection. Yet, for each type of fault class, there exists a subset of features with common values and correlation which are specific for that class. This unique subset of features forms the artifact.

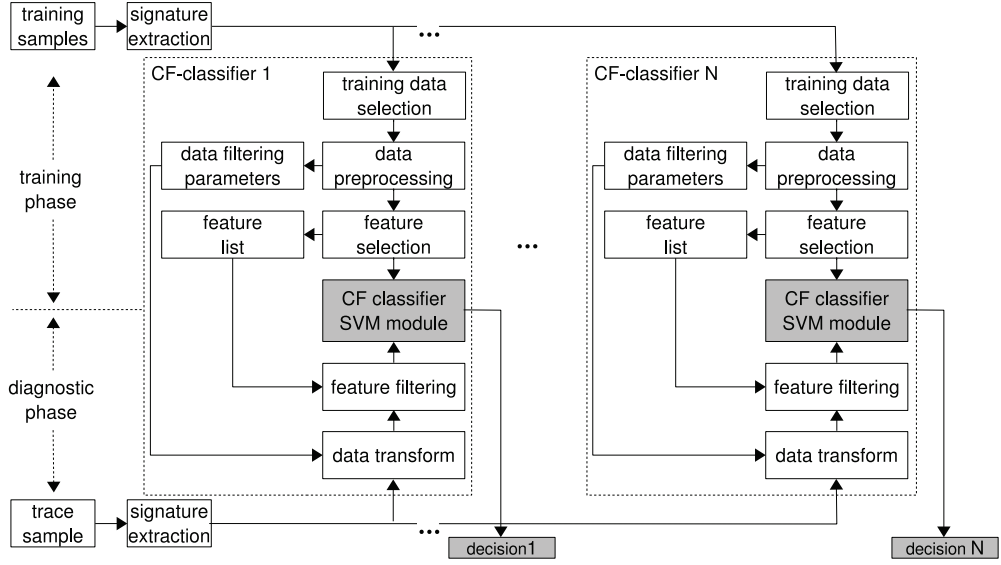

Fig. 2. CFD classifier design for the IACD system.

### C. Data Pre-Processing

The raw feature vectors are further processed before being used for classifier training. This step improves the overall classification accuracy by enhancing data coherency and consistency within the classes. First, categorical attributes such as the class labels $cf_0$ and $cf_j$ are converted to numeric data as in

$$y_i \equiv cf_0 \equiv -1, \quad (4a)$$
$$y_i \equiv cf_j \equiv +1 \text{ for } j = 1, ..., p. \quad (4b)$$

Then the data is shifted and linearly re-scaled along each feature to fit in the range 0-1. Data re-scaling avoids numerical difficulties and avoids features with greater numeric range dominating the smaller.

### D. Feature Selection

Although the signature format is identical in every sample, only a particular subset of features contribute to the artifact. We use a simple automated feature selection method to select the best suitable feature subset for a particular classifier. Our proposed feature selection method is shown in Figure 3. This algorithm, similar to work discussed by Xing et al. [21] and Das et al. [22], follows a hybrid approach (between filter and wrapper) for automatically isolating the best feature subsets. We first use a filter, Student's t-test (two sample t-test) (implemented similar to [23]) to assess the significance of every feature for separating the two classes. Next, the features sorted in the order of significance are cross-validated by incrementing the number of features selected for each class (wrapper technique) against test samples to identify the best number of features required for each classifier. The feature selection process reduces the $m$-dimensional feature vector in (1) to $q$-dimensions, where the combination of $q$ features create the artifact. Note that further fundamental analyses of the relationship between selected features and client faults can be facilitated by, and in turn aid in refining, the feature selection process.

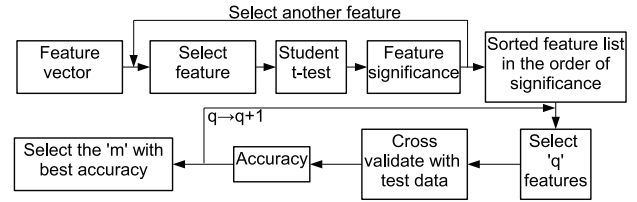

Fig. 3. Hybrid feature selection technique for isolating the best feature subset of the artifact.

## IV. FAULT DIAGNOSIS

Once the training phase is completed, each classifier module contains (i) filtering parameters (scale factors) derived during pre-processing, (ii) a feature list chosen during feature selection, and (iii) a classifier created using SVMs. During the diagnostic phase, packet traces are captured using the trace collection application at the undiagnosed client. Then the traces are sent through the IACD system which feeds the traces into the CF-classifier network. Each CF-classifier then identifies if the particular fault exists in the trace. The collective output of $n$ CF-classifiers determine the faults that exists in the client. However, at this stage, the system is not capable of diagnosing unknown faults in a trace.

## V. PERFORMANCE EVALUATION

### A. Network Emulation and Experiment Criteria

As a proof of concept, data was collected in a network test bed which emulated an access link, client computer and the access server as shown in Figure 4. The client and server used Linux 2.6.32 systems (with Ubuntu distribution), capable of running multiple TCP variants. The access link was emulated using a network emulator, dummynet [24] on FreeBSD 7.3. Each box was connected using full-duplex cat5e ethernet. The dummynet emulated a full-duplex wired access link with a 80 Mb/s bandwidth, 10 ms delay with no packet losses and no packet reordering. Faulty links conditions were created by creating packet losses (from 1% up to 10%) and increased delays (from 15ms up to 100ms). Both the server, dummynet box, and the healthy client had an optimized protocol stack to avoid bottlenecks. Client faults were emulated using the Linux TCP parameter configurations. Then the traces were captured using the technique discussed previously.

The performance of the IACD system was analyzed with a network of four CF-classifier modules. We emulated the disabled Selective Acknowledgement (SACK) option (CF-Classifier 1) and the disabled Duplicate Selective Acknowledgement (D-SACK) option (CF-Classifier 2), which have been found to cause performance issues in the high bandwidth connections [25]–[28]. Also, Socket buffer limitations, another common and hard to diagnose performance bottleneck [6], [29] were emulated by creating insufficient read buffers (CF-Classifier 3) and write buffers (CF-Classifier 4) at the client as two separate cases. Multiple, simultaneous client faults were emulated by creating both read and write socket buffer limitations at the same time. All buffer limitations were emulated using three buffer levels to collect traces from a range of possible scenarios.

For training data, both the server and client were limited to run TCP-CUBIC [30] with only 11 traces per each fault class being collected to re-create the worst case practical limitations. To analyze the system performance, we collected four testing data sets as follows: (i) the same data set used in training and, separately collected sets with (ii) TCP-CUBIC client, (iii) TCP-BIC [31] client, (iv) TCP-NewReno [32] client. The data sets (iii) and (iv) were collected with other TCP variants to evaluate the TCP agnostic properties of the system.

### B. Diagnostic Performance

Figure 5 shows the diagnostic accuracy of the CFD classifier, which considers the collective decision taken using the output of the CF-classifier network. When tested with the CUBIC training and testing data sets, the system was capable of diagnosing the client's disabled SACK option, disabled D-SACK option, read buffer limitation and write buffer limitations with high accuracy. Similarly, when tested with TCP-BIC and TCP-NewReno, variants not used during the training phase, the four client faults were diagnosed with 100% accuracy. These results suggest that proposed CFD classifier design may be independent of TCP variant.

The healthy clients were identified with a 94.81% and 93.5% accuracy during the first two tests of TCP-CUBIC train and test data sets. When samples from healthy clients with TCP-BIC and TCP-NewReno were tested, the detection accuracies were at 92.10% and 91.71%, marginally lower than the other cases. This is due to the slightly higher tendency of obtaining a false positive in at least one of the CF-classifiers by healthy clients' traces compared to other samples. When presented with traces taken from clients with simultaneous read and write buffer deficiencies, CF-classifier 3 and CF-classifier 4 were capable of independently identifying the faults from the trace. This capability led to a collective diagnostic accuracy of 96.97%, 96.90% and 100% for CUBIC, BIC and NewReno data sets, respectively.

## VI. SYSTEM CHARACTERISTICS AND COMPARISON WITH THE SIMILAR WORK

For the root cause diagnosis of client performance problems, the proposed IACD system offers many advantages over the other available trace inference methods.

- The system offers a fully-automated, comprehensive framework which is extendible to diagnose a diverse range of faults, contrary to the limited capabilities of offered by tools that uses TCP traces for information gathering and measurement purposes [9], [11], [33].
- Diagnostic capability of the system evolves with the diversity of the fault signature databases, instead

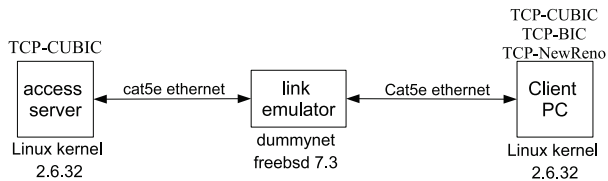

Fig. 4. Network testbed for training and testing data collection

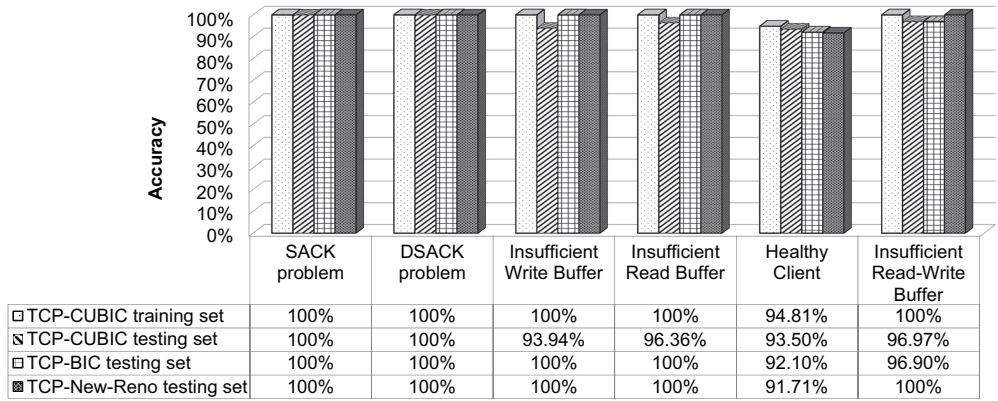

Fig. 5. Diagnostic accuracy of the CFD classifier, derived from the collective output of the CF-classifier network

- of the inference algorithm. Users can collaborate to create common signature repositories, encompassing a wide range of faults, networks, and client platforms. Most rule based systems are limited to a specific set of faults and lack the generality to operate effectively in a dynamic environment [7], [8], [10], [34].
- The system relies solely on packet traces collected at end-points and can be implemented as an application. This provides flexibility for the operator to deploy the IACD system at any desired network location. Popular client diagnostic solutions, mainly based on Web100 TCP kernel instrumentation require changes to the kernel and the system it self [10], [35], [36].
- End-user systems can be diagnosed without remotely accessing or physically logging on to the systems; a capability unavailable in many network diagnostic tools. Most machine learning based solution such as "NEVERMIND" [37], "pinpoint" [38], "Netprints" [33] require information such as user requests, event logs, system calls or private network traffic which demands privileged access.
- The proposed technique, contrary to many other similar work, avoids both the idiosyncrasies of individual TCP implementation and the usage of TCP flags as an information source as they sometimes are misleading [7]–[9], [39]. Instead, the connections are characterized using per-connection statistics without relying on the negotiated flags and independent of the TCP variant.
- Although the system is designed to diagnose client terminals accessing the cloud, the same system can be used for diagnosing internal nodes of the cloud by deploying a trace collection module in a neighboring node and training with suitable data.

## VII. CONCLUSION AND FUTURE DIRECTIONS

We have proposed and evaluated the IACD system, an automated client diagnostic system for private cloud environments. IACD system uses an intelligent inference based approach of TCP packet traces to identify artifacts created by client faults. The CFD classifier performs a complex multi class classification of client faults using a parallel network of SVM-based CF-classifiers. The modular design of the CFD classifier offers extendibility to diagnose new faults by training CF-classifier modules independently.

We evaluated the system by diagnosing four types of common client problems with various TCP implementations. Additionally, we analyzed system performance in the absence of any client faults and also in the case of multiple simultaneous faults. Our results show that, with a small number of training samples, CF-classifier modules collectively produce high diagnostic accuracy in all tested scenarios, including clients with different faults, TCP variants, default clients and multiple faults.

To our knowledge, the IACD system is the first framework for automating the client diagnosis with TCP packet trace-based fault signatures and SVM-based learning. The work presented in the paper only serves as a feasibility study to explore the system capabilities in a limited network environments. This work provides the foundation to extend the system to a more complex cloud computing environments such as public and hybrid clouds with thousands of users, diverse client platforms and extremely complex traffic patterns.